\documentstyle[axodraw]{article}          %% 
\title{\bf Nucleon instability in a 
supersymmetric $\bf SU(3)_c\times SU(3)_L\times U(1)$ model}

\author{{\bf Hoang Ngoc Long}$^{\rm a}$
and
{\bf Palash B. Pal}$^{\rm b}$
\\ 
\normalsize 
a) %Institute of Physics, NCNST, 
Department of Physics,
Chuo University, Kasuga, Bunkyo-ku, Tokyo 112, Japan\\
\normalsize 
Email:  hnlong@phys.chuo-u.ac.jp\\
\normalsize 
b) Theory Group, Saha Institute of Nuclear Physics, 
Calcutta 700064, India\\ 
\normalsize 
Email: pbpal@tnp.saha.ernet.in
}

\date{}

\begin{document}
\twocolumn[

\maketitle

%%%%%%%%%%%%%%%%%%%%%%%%%
\begin{quotation}

We construct the supersymmetric version of a model based on the gauge
group $\rm SU(3)_c\times SU(3)_L\times U(1)$.  We discuss the
mechanism of baryon number violation which induces nucleon decay, and
derive bounds on the relevant couplings. We point out a new mechanism
for nucleon decay which can be present in R-violating MSSM as well.

\bigskip\bigskip

PACS number(s): 11.30.Fs, 12.60.Jv, 12.60.Cn
%(baryon, lepton number, supersymmetric model, extension
% of electroweak gauge sector) 

\end{quotation}
%%%%%%%%%%%%%%%%%%%%%%%%%
\vspace*{5mm}
]

The standard model of particle interactions, based on the gauge
group $\rm SU(3)_c \times SU(2)_L \times U(1)_Y$ is very
successful experimentally. But it does not answer or address some
important theoretical questions. One of these, for example, is
the question of the number of generations of fermions. At
present, we know of three generations, but the standard model
does not explain why this number has to be three.

This question obtains a natural answer in an interesting
extension \cite{PiPl92,Fra92} of the standard model, based on the
gauge group $\rm SU(3)_c \times SU(3)_L \times U(1)_N$. In these
theories, the fermion spectrum is extended, by including new
quark-type fields, in such a way that chiral anomalies do not
cancel in any single generation of fermion fields. The
generations, on the other hand, are not exact replicas of each
other, and all gauge anomalies cancel when all three generations
are taken into account. Thus, this model requires the number of
generations to be 3, or any multiple of 3. Another interesting
feature of this model is that the Peccei-Quinn symmetry,
necessary to solve the strong-CP problem, follows naturally
\cite{Pal} from the particle content in these models.  The aim of
this paper is to construct a supersymmetric version of this model
and discuss what sort of baryon-number violating processes arise
from such an extension.

We start with introducing the chiral superfields of this
model. We start with the superfields that contain the quarks and
leptons. We write only the left-handed fields throughout this
paper, and omit any subscript $L$ for them.
	\begin{eqnarray}
\Psi_a &:& (1,3,0) \nonumber\\*
Q_1 &:& (3,3,{2/3}) \nonumber\\ 
Q_i &:& (3,\bar 3,-\,{1/3}) \nonumber\\ 
U^c_a &:& (\bar 3,1, -\, {2/3}) \nonumber\\ 
D^c_a &:& (\bar 3,1, {1/3}) \nonumber\\
T^c_1 &:& (\bar 3,1, -\, {5/3}) \nonumber\\* 
B^c_i &:& (\bar 3,1, {4/3}) \,.
\label{qandl}
	\end{eqnarray}
Here, $a$ is a generation index that runs from 1 to 3. The other
generation index, $i$, runs only from 2 to 3. Thus, the field
content of the first generation is different from that in the
other two. 

The electric charge generator is defined as
	\begin{eqnarray}
Q_{\rm em} = \lambda_{3L} + \sqrt 3 \lambda_{8L} + N \,,
	\end{eqnarray}
where $\lambda_{3L}$ and $\lambda_{8L}$ are the two diagonal 
generators of $\rm SU(3)_L$. In the fundamental representation, 
these are given by
	\begin{eqnarray}
\lambda_{3L} &=& {1\over 2} \; {\rm diag} \left( 1,-1,0
\right) \,, \nonumber\\*
\lambda_{8L} &=&  {1 \over 2\sqrt 3} \; {\rm diag} \left( 1,1,-2
\right) \,,
	\end{eqnarray}
and $N$ is the generator of $\rm U(1)_N$ whose quantum numbers
are given in Eq.~(\ref{qandl}). Using these formulas, we see that
the charges of the members of $\Psi_a$ should be $+1,0,-1$, and
the fermionic components of these fields can be identified with
the left-chiral projection of the antilepton, the neutrino and
the lepton in a given generation. The right-handed neutrino
field, unknown so far from experiments, have not been introduced
in this model, so there is no left-handed antineutrino field. 

Looking now at the componets of $Q_1$, we find that the two
lower components have the charges of the $u$ and the $d$ quarks,
and the fermionic fields for these components are identified with
these quarks. The uppermost component has now a charge $5/3$,
which is one of the exotic quarks in this model. Let us call it
$T_1$. Its right-handed counterpart is the antiparticle of the
field $T^c_1$ given in Eq.\ (\ref{qandl}). Similarly, in
$Q_i$, we have the fields known in the standard model, plus
extra quark fields with charge $-4/3$, to be called $B_i$. The
right-handed counterparts of these fields are conjugates of
$B^c_i$ which appear in Eq.~(\ref{qandl}), where we also have
the conjugates of the right-handed counterparts of the up-type
and down-type quark fields present in the standard model.

So far, we have used only the quark and lepton fields present in the
non-supersymmetric version of the model, and their
superpartners. For the Higgs superfields, however, we must do
something more. In the non-supersymmetric version, it was argued
\cite{FHPP} that the following Higgs multiplets can break the
symmetry and give reasonable masses to the quarks and the
leptons: 
	\begin{eqnarray}
\chi : (1,3,-1) &,&
\rho : (1,3,1) \nonumber\\*
\eta : (1,3,0) &,&
S : (1,\bar 6,0) \,.
\label{higgs}
	\end{eqnarray}
In a supersymmetric version, the superpartners of these fields
will give rise to chiral anomalies. Thus, to cancel them, we need
other fields. The most obvious choice is a set of fields which
are exactly in the complex conjugate representation of the gauge
group. Let us call these fields $\chi^c$, $\rho^c$, $\eta^c$ and
$S^c$. Thus, for example, $\chi^c$ would transform like $(1,\bar
3,1)$, and so on.

With this field content, we now write down the superpotential of
the model. This is:
	\begin{eqnarray}
W &=& h^{(1)} Q_1 T^c_1 \chi^c + h^{(2)}_{ij} Q_i B^c_j \chi \nonumber\\* 
&& + h^{(3)}_{a} Q_1 D^c_a \rho^c + h^{(4)}_{ia} Q_i 
U^c_a \rho \nonumber\\* 
&& + h^{(5)}_{a} Q_1 U^c_a \eta^c + h^{(6)}_{ia} Q_i D^c_a \eta
\nonumber\\*  
&& + h^{(7)}_{ab} \Psi_a \Psi_b \eta + h^{(8)}_{ab} \Psi_a
\Psi_b S^c \nonumber\\
&& + \lambda_{abc} \Psi_a \Psi_b \Psi_c + \lambda'_{aib} \Psi_a Q_i D^c_b + 
\lambda''_{abc} U^c_a D^c_b D^c_c \nonumber\\* 
&& + \kappa_{ai} T^c_1 D^c_a B^c_i + \mu_a \Psi_a \eta^c \nonumber\\
&& + f_1 \eta \eta S^c + f'_1 \eta^c \eta^c S 
+ f_2 \chi\rho\eta + f'_2 \chi^c\rho^c\eta^c \nonumber\\*
&& + f_3 \chi\rho S^c + f'_3 \chi^c \rho^c S \nonumber\\*
&& + m_\chi \chi\chi^c + m_\rho \rho\rho^c + m_\eta \eta\eta^c + m_S
SS^c \,.
\label{supot}
 	\end{eqnarray}
All generation indices are assumed to be summed over the
respective range of values. The terms $h^{(1)}$ through $h^{(8)}$
give the Yukawa couplings present in the non-supersymmetric
version of the model. The terms with the couplings $f$, $f'$ and
the bilinear terms appearing after them in Eq.\ (\ref{supot}) involve
the Higgs-type 
fields only. Finally, note that the couplings $\lambda''$ and
$\kappa$ involve baryon number violation. Of these, $\kappa$
involves more than one exotic quark fields, and cannot affect
processes involving the usual quarks at the tree level. Thus, in
order to look at baryon number violating processes, we need to
look at the couplings $\lambda''$. Such couplings are present in
the supersymmetric standard model as well, and, like in that
case, the couplings satisfy
	\begin{eqnarray}
\lambda''_{abc} = - \, \lambda''_{acb}
\label{antisymm''}
	\end{eqnarray}
because of the exchange symmetry of the $D^c$-fields. The same
argument dictates that the couplings with the term
$\Psi_a\Psi_b\Psi_c$ is totally antisymmetric in the generation
indices. So, for three generations, there is only one such
independent coupling, which we will denote simply by $\lambda$.

%%%%%%%%%%%%%%%%%%%%%%%%%
\begin{figure}
\begin{center}
\begin{picture}(150,100)(0,0)
\SetWidth{1.2}
\ArrowLine(10,10)(50,50)
\Text(25,30)[r]{$u^c$}
\ArrowLine(10,90)(50,50)
\Text(25,70)[r]{$d^c$}
\DashArrowLine(100,50)(50,50){3}
\Text(75,58)[b]{$\widetilde d^c_j$}
\ArrowLine(100,50)(140,90)
\Text(125,70)[l]{$q_i$}
\ArrowLine(100,50)(140,10)
\Text(125,30)[l]{$\psi_a$}
\end{picture}
\end{center}
\caption[]{Proton decay mediated by $\lambda''$ and $\lambda'$
couplings.}\label{l''l'} 
\end{figure}
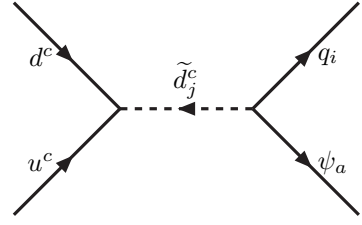
%%%%%%%%%%%%%%%%%%%%%%%%%
It is worthwhile to note that, unlike the minimal supersymmetric standard
model (MSSM), one cannot impose an $R$-parity to eliminate all baryon
and lepton number violation in this model. The reason is that there
are lepton number violating interactions even in the gauge sector of
this model. Baryon number violation can be eliminated, for example, by
introducing a discrete $Z_2$ symmetry under which all the superfields
presented in Eq.\ (\ref{qandl}) change sign, whereas those presented
in Eq.\ (\ref{higgs}) do not. But we take the most general
superpotential allowed by gauge symmetry and supersymmetry, which is
the one in Eq.\ (\ref{supot}), and examine its consequences for
nucleon decay.

In order to have proton decay, one needs lepton number violation
in addition to baryon number violation. In this model, lepton
number violation has two different sources. Even in the
non-supersymmetric version, lepton number violation was present
due to $\rm SU(3)_L$ gauge interactions, since the charged
antilepton is put in the same multiplet as the charged lepton and
the neutrino. In the supersymmetrized version, additional lepton
number violation comes in through variuos Yukawa terms which can
be derived from the superpotential. We will see that these are
the terms which give leading  contributions to proton decay in
this model.

One such contribution is shown in Fig.~\ref{l''l'}, which is
mediated by a squark of charge $-1/3$. Such contributions to
proton decay exists even in the MSSM \cite{rpar,review}. There is,
however, one distinctive feature 
here. Notice that the coupling $\lambda'$ involves the multiplet
$Q_i$, where $i$ can take only the values 2 or 3. Thus, 
the outcoming quark can belong only to the
second or the third generation, barring contributions coming from
intergenerational mixing. So the dominant decay mode arising out of
this diagram would be
	\begin{eqnarray}
p \to K^+ \nu_a \,,
	\end{eqnarray}
where the neutrino can be from any generation, depending on the
relative strengths of the different couplings $\lambda'_{aib}$. The
effective coupling for this decay will be given by
	\begin{eqnarray}
G_{\rlap/B} \simeq 
{\lambda''_{11j} \lambda'_{a2j} \over m^2_{\widetilde d_j}} \,,
	\end{eqnarray}
where the generation index $j$, as mentioned before, can take
only the values 2 and 3 in view of the antisymmetry property of
the couplings $\lambda''$ mentioned in
Eq.~(\ref{antisymm''}). This will lead to a lifetime of 
	\begin{eqnarray}
\tau_p \simeq \left( m_p^5 G_{\rlap/B}^2 \right)^{-1} \,. 
	\end{eqnarray}
The experimental lower bound on these modes is about $10^{32}$
years. Using that, we obtain
	\begin{eqnarray}
\lambda''_{11j} \lambda'_{a2j} < 10^{-24} \,,
	\end{eqnarray}
assuming the superpartner masses in the range of 1\,TeV.

If we consider the effects of intergenerational mixing, we can
extend the bounds \cite{SmVi96} to any product of the couplings
of the form $\lambda''\lambda'$. Of course, intergenerational
mixings will suppress decays like $p\to \pi^+\nu$, and therefore
the bounds will be a little weaker.

%%%%%%%%%%%%%%%%%%%%%%%%%
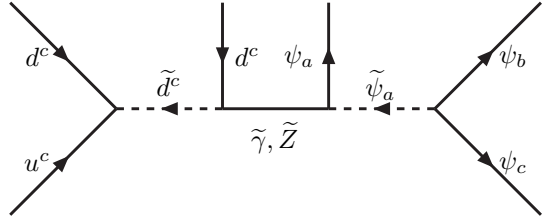
\begin{figure}
\begin{center}
\begin{picture}(220,100)(0,0)
\SetWidth{1.1}
\ArrowLine(10,10)(50,50)
\Text(25,30)[r]{$u^c$}
\ArrowLine(10,90)(50,50)
\Text(25,70)[r]{$d^c$}
\DashArrowLine(90,50)(50,50){3}
\Text(70,55)[b]{$\widetilde d^c$}
\ArrowLine(90,90)(90,50)
\Text(95,70)[l]{$d^c$}
\Line(90,50)(130,50)
\Text(110,45)[t]{$\widetilde \gamma, \widetilde Z$}
\ArrowLine(130,50)(130,90)
\Text(125,70)[r]{$\psi_a$}
\DashArrowLine(170,50)(130,50){3}
\Text(150,55)[b]{$\widetilde\psi_a$}
\ArrowLine(170,50)(210,90)
\Text(195,70)[l]{$\psi_b$}
\ArrowLine(170,50)(210,10)
\Text(195,30)[l]{$\psi_c$}
\end{picture}
\end{center}
\caption[]{Proton decay mediated by $\lambda''$ and $\lambda$
couplings.}\label{l''l}
\end{figure}
%%%%%%%%%%%%%%%%%%%%%%%%%
Let us now look at a different mechanism for nucleon decay. The
relevant diagram has been shown in Fig.~\ref{l''l}. Here, one
utilizes the couplings $\lambda''$ and $\lambda$. Again, the
vertex at the right end of the figure exists in the MSSM
in the form of the coupling $LLE^c$ and so this kind of diagram exists
even in the MSSM \cite{SmVi??}. 
But the distinctive feature here is
that there is only one coupling $\lambda$, which connects three
different generations of lepton fields. Thus, the vertex at the
right end of the diagram must have fields from all three
different generations. Let us also assume that, to a first
approximation, the couplings of the gaugino fields are flavor
diagonal. In this case, the three outgoing leptonic
fields belong to three different generations. Since the $\tau$-lepton
is heavier than the nucleon, this means that the charged leptons
available in the decay product must be $\mu^+e^-$ or
$\mu^-e^+$. In other words, we obtain the following decay modes
at the quark level:
	\begin{eqnarray}
u^c d^c s^c \to \mu^\pm e^\mp \nu_\tau \,,
	\end{eqnarray}
taking into account that the couplings $\lambda''$ must be
antisymmetric in the down-type quark indices.
For the proton, it implies the decay mode
	\begin{eqnarray}
p \to K^+ \mu^\pm e^\mp \bar\nu_\tau \,.
	\end{eqnarray}
Of course, we can also have $p \to \pi^+ \mu^\pm e^\mp
\bar\nu_\tau$ etc, but those will be suppressed by
intergenerational mixings.

The effective operator here is a six-fermion one, with the
effective coupling
	\begin{eqnarray}
G'_{\rlap/B} \simeq 
{g^2 \lambda''_{112} \lambda \over M_{\widetilde Z} m^2_{\widetilde d_i}
m^2_{\widetilde \psi_i}} \,. 
	\end{eqnarray}
This will give a lifetime
	\begin{eqnarray}
\tau_p \simeq \left( m_p^{11} G'_{\rlap/B} {}^2 \right)^{-1}\,. 
	\end{eqnarray}
There are no direct experimental limits on the specific decay
modes obtained here. But if we take the lower limit of $10^{31}$ years as
a benchmark value, we obtain 
	\begin{eqnarray}
\lambda''_{112} \lambda < 10^{-16} \,,
	\end{eqnarray}
assuming the masses of the supersymmetric particles to be of
order 1\,TeV.

In addition, as we mentioned earlier, lepton number violation can come
from $\rm SU(3)_L$ gauge couplings. However, these cannot induce
proton decay at the tree level. The reason is that in the quark
sector, such couplings involve exotic quarks, which will have to be
much heavier than the proton.

Notice that we have dealt with a specific version of $\rm
SU(3)_c\times SU(3)_L\times U(1)$ models. One can apply similar
ideas to a different model containing right-handed
neutrinos~\cite{flt}. The consequences are similar, so we do not
discuss it in detail.

We thank A. Smirnov, F. Vissani and G. Bhattacharyya for discussions.
This work was written up during a workshop on Highlights of
Astroparticle Physics organized by the International Centre for
Theoretical Physics, and we thank the organizers for hospitality.

\end{document}